\documentclass[]{aa}
\usepackage{graphicx,txfonts,amssymb,natbib}
\renewcommand{\d}{\mathrm{d}}
\authorrunning{C. Fedeli et al.}
\titlerunning
  {Strong lensing statistics and the power spectrum normalisation}

\begin{document}

\title
  {Strong lensing statistics and the power spectrum normalisation}

\author{C. Fedeli\inst{1,2,3,4}\thanks{E-mail: cosimo.fedeli@unibo.it}, M. Bartelmann\inst{1}, M. Meneghetti\inst{3,4} \and L. Moscardini\inst{2,4}}
 
 \institute{$^1$ Zentrum f\"ur Astronomie, ITA, Universit\"at Heidelberg,
     Albert-\"Uberle-Str. 2, 69120 Heidelberg, Germany\\$^2$
     Dipartimento di Astronomia, Universit\`a di Bologna,
     Via Ranzani 1, 40127 Bologna, Italy\\$^3$ INAF-Osservatorio
     Astronomico di Bologna, Via Ranzani 1, 40127 Bologna, Italy\\$^4$
     INFN, Sezione di Bologna, viale Berti Pichat 6/2, I-40127 Bologna, Italy}
 
\date{\emph{Astronomy \& Astrophysics, submitted}}

\abstract{We use semi-analytic modelling of the galaxy-cluster population and its strong lensing efficiency to explore how the expected abundance of large gravitational arcs on the sky depends on $\sigma_8$. Our models take all effects into account that have been shown to affect strong cluster lensing substantially, in particular cluster asymmetry, substructure, merging, and variations in the central density concentrations. We show that the optical depth for long and thin arcs increases by approximately one order of magnitude when $\sigma_8$ increases from $0.7$ to $0.9$, owing to a constructive combination of several effects. Models with high $\sigma_8$ are also several orders of magnitude more efficient in producing arcs at intermediate and high redshifts. Finally, we use realistic source number counts to quantitatively predict the total number of arcs brighter than several magnitude limits in the R and I bands. We confirm that, while $\sigma_8\sim0.9$ may come close to the known abundance of arcs, even $\sigma_8\sim0.8$ falls short by almost an order of magnitude in reproducing known counts. We conclude that, should $\sigma_8\sim0.8$ be confirmed, we would fail to understand the strong-lensing efficiency of the galaxy cluster population, and in particular the abundance of arcs in high-redshift clusters. We argue that early-dark energy or non-Gaussian density fluctuations may indicate one way out of this problem.}


\maketitle

\section{Introduction}

After the 3-year WMAP data release \citep{SP07.1}, the best-fit value for the normalisation of the density-fluctuation power spectrum was lowered from $\sigma_8 \sim 0.9$ to $\sigma_8 \sim 0.75$. This puts many of the cosmological tests based on structure formation under stress, such as the number counts of galaxy clusters and its evolution \citep{EV08.1,RI07.1}, the large-scale structure probed via weak gravitational lensing, and large optical surveys that tend to favour a value of $\sigma_8 \sim 0.9$ as well \citep{HO06.1}. The value of $\sigma_8$ is perhaps the least well-known of the main cosmological parameters today. The controversy is highlighted by the fact that largely discrepant values of $\sigma_8$ are being published \citep{RE06.1,EV08.1}

At the same time, there are signs of convergence on a normalisation which may be compatible with most measurements of the amplitude of structures large enough for linear evolution to dominate. At $\sigma_8\sim0.8$, CMB and gravitational-lensing measurements may meet after a slight increase in the mean redshift of the background sources used to identify the weak-lensing signal \citep{FU08.1}. Some analyses of the X-ray cluster population also seem to find this value acceptable \citep{RE06.1}.

Suppose, then, that $\sigma_8\sim0.8$. We argue here that this exacerbates a problem with non-linear structure growth on the cluster scale to a level which seems serious despite considerable uncertainties. Specifically, we shall address the statistics of gravitational arcs in galaxy clusters to show that none of the many possible explanations suggested in the past decade suffices to bring theoretical expectations into agreement even with the admittedly sparse observations.

We summarise the situation in Sect.~2 and compile some analytical results concerning model universes with different normalisation of the power spectrum in Sect.~3, in order to gain quantitative insight into the situation. In Sect.~4, we describe the semi-analytic modelling of lensing by the galaxy-cluster population. Section 4 summarises the results, and Sect.~5 presents our conclusions. There, we also discuss how our results comply with the new WMAP-5 data release \citep{DU08.1,KO08.1} which was published after this work was completed.

\section{Is there an arc-statistics problem?}

Comparing the observed abundance of gravitational arcs to the efficiency of numerically simulated galaxy clusters for strongly lensing distant galaxies, \cite{BA98.2} claimed that about an order of magnitude fewer arcs are expected in the $\Lambda$CDM cosmology than are actually observed. This was called the arc-statistics problem. Numerous attempts were carried out to see how the problem could be solved. A corrugation of the lensing potential by individual cluster galaxies turned out to have a slight net effect on large arcs because the increased length of the caustic curves was counteracted by splitting arcs into shorter pieces \citep{FL00.1,ME00.1}.

Cluster asymmetry, however, was identified as crucial for the abundant formation of arcs \citep{BA95.1,MO01.1,ME07.1}. Simplified analytic cluster models were unable to reproduce the numerical results \citep{CO99.2,KA00.1} for two reasons; first, they did not account for the dependence of cluster concentrations on the cosmological constant; and second, simple elliptical cluster models were also found inadequate to quantitatively explain the arc-formation efficiency in numerically simulated models \citep{ME03.1}. Lensing properties of numerically simulated galaxy clusters were found to agree well with those of comparable, real clusters \citep{HO05.1}.

Based on numerical simulations, \cite{WA04.2} found that the probability for high magnifications along light rays propagating to us from sources in the distant Universe depends steeply on the source redshift distribution. Based on this result, they concluded that there is in fact no arc-statistics problem if the realistically distant tail of the source distribution is taken into account. However, their identification of highly magnified light bundles with strongly distorted arcs was questioned by \cite{LI05.1}, who showed that at least numerically simulated clusters produce a class of highly magnified, but weakly distorted images. In particular, numerical simulations show that the most likely length-to-width ratio for arcs magnified by a factor $\ge 10$ is typically $\sim 3$ (see also Figure~\ref{fig:conditional} below).
High magnification probability does therefore not imply a frequent occurrence of strongly distorted arcs. Identifying arcs in simulations, the redshift dependence found by \cite{WA04.2} was qualitatively confirmed, but quantitatively found to be substantially weaker \citep{LI05.1,FE06.1}.

The arc-statistics problem was again questioned by \cite{DA03.1}, who used numerical simulations to confirm the overall optical depth found by \cite{BA98.2}, but found agreement between the observed and expected numbers of arcs because they estimated a lower abundance of observed arcs and a higher number density of background sources. They took the source-redshift distribution into account, but found it weaker than \cite{WA04.2} had claimed.

The problem did not disappear, however, because it was observationally found that in particular the number of arcs in distant clusters is considerably higher than naively expected \citep{TH01.1,GL03.1,ZA03.1}. \cite{OG03.1} studied the effect of central concentration and triaxiality of cluster-sized halos on their strong-lensing ability, finding that triaxial clusters with sufficiently steep central density profile can come close to the observed results. While this effect is undoubtedly present, it is also included in numerically simulated clusters and does thus not attenuate the apparent discrepancy between the arc abundances observed in the sky and produced in realistically simulated cluster populations.

So, is there an arc-statistics problem? Based on the preceding discussion, the question still seems undecided. However, it is important to note that up to this point, all simulations and calculations were done assuming $0.9\le\sigma_8\le1.2$. \cite{LI06.1} pointed out that the problem is substantially aggravated if $\sigma_8$ is lowered to the value preferred by the third-year WMAP data analysis, $\sigma_8=0.74$ \citep{SP07.1}. \cite{LI06.1} found that the expected abundance of large arcs on the sky drops very steeply if $\sigma_8$ is lowered, reflecting the exponential decrease of the abundance of massive clusters with $\sigma_8^2$.

Motivated by this earlier study, in which numerical simulations with two different values of $\sigma_8$ were compared, we shall investigate in this paper how the number of gravitational arcs predicted in a standard cosmological model changes with $\sigma_8$, and how the results compare with the observed statistics. Such a study is possible only because we can replace numerical simulations by semi-analytic calculations using a novel algorithm developed by \cite{FE06.1}. The method captures two ingredients crucially important for strong cluster lensing: cluster asymmetries and cluster mergers \citep{TO04.1}, which were found to have transient, but strong effects on arc cross sections \citep{FE06.1}. We also properly account for a realistic source redshift distribution and the luminosity function of background sources. The contribution of cD galaxies is ignored. \cite{ME03.2} showed that the cross section for giant arcs of numerical clusters is increased by $\lesssim 50\%$ when a realistic cD model is included, and this does not significantly alter our conclusions. In addition, the same paper shows that the boosting effect of a massive central galaxy is much reduced when the host model cluster is asymmetric, as those used here. We also neglect the effects of gas physics, which are potentially important in cluster cores. \cite{PU05.1} have shown that, depending on the physical effects included into the modelling, the lensing cross section of individual clusters can be increased by $\lesssim 100\%$ due to the presence of baryonic matter. \cite{WA08.1} implemented a specific model describing baryon cooling and galaxy formation in the core of cluster-sized dark matter halos, finding an increase in the production of arcs of $\sim 25\%$. \cite{HI07.2} (see also \citealt{HI07.1}) performed a very similar study on the effect of the stellar component on optical depths for image splitting in the Millennium Simulation, finding results compatible with earlier analyses \citep{ME00.1}. Although baryonic physics may increase strong-lensing cross sections, it is not expected to bridge the order-of-magnitude gap between theory and expectations, although the uncertainties in the modelling are admittedly substantial.

We shall consider five different cosmological models. The energy density content and the Hubble constant are taken from the WMAP-3 data combined with the SDSS observations \citep{SP07.1}, and kept fixed throughout the paper. They are $\Omega_{\mathrm{m},0} = 0.265$, $\Omega_{\Lambda,0} = 0.735$ and $h = 0.71$. The values of $\sigma_8$ are chosen differently for the five models and are $0.7$, $0.75$, $0.8$, $0.85$ and $0.9$ respectively. This choice allows us to cover the complete range of values from the third-year WMAP data to normalisation values derived from galaxy-cluster counts.
Typical degeneracies among different cosmological parameters imply that changes in $\sigma_8$ would cause $\Omega_{\mathrm{m},0}$ and possibly other parameters to change, depending on the data set underlying the parameter determination. However, since the main purpose of this work is to isolate the impact of the power-spectrum normalization on arc statistics, we choose to keep everything fixed except $\sigma_8$. We shall discuss the effect of relaxing this assumption in the last section of the paper.

\section{Expectations}\label{sct:exp}

We need theoretical predictions of the arc abundance which are at the same time as precise as possible and fast to achieve because they need to be carried out in many cosmological models. We first try to gain some insight into the various contributions to the arc optical depth, its dependence on and its variation with the normalisation of the power spectrum.

We start with the internal structure of dark matter halos that, as described in a variety of studies \citep{CO96.1,NA97.1,JI00.1,BU01.1,EK01.1}, depends on the complete halo formation history. Assuming that the density profile of cluster-sized dark matter halos is of NFW form \citep{NA95.3,NA96.1,NA97.1}, we first explore how the concentration of the profile depends on $\sigma_8$. The concentration of a dark matter halo is the ratio of the virial radius to the scale radius $r_\mathrm{s}$ of the density profile. Here we define the virial radius as the radius of the sphere inside which the average density of the halo is 200 times the critical density of the Universe at the given redshift. This radius separates well the internal, relaxed part of isolated galaxy clusters in numerical simulations from the external, infall part \citep{EK98.1}. We expect that higher $\sigma_8$ allows earlier structure formation, such that halos form from a higher background density and have more time to relax, causing higher concentrations.

\begin{figure}[t]
  \includegraphics[width=0.5\hsize]{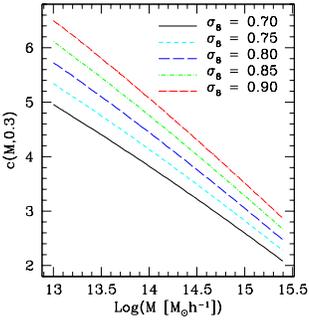}\hfill
  \includegraphics[width=0.5\hsize]{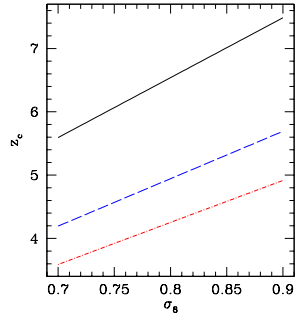}
\caption{\textit{Left panel}: Concentration of dark matter halos according to \cite{EK01.1} as a function of virial mass at fixed redshift, $z=0.3$. Five different values for $\sigma_8$ are considered, as labelled in the plot. \textit{Right panel}: Collapse redshift according to \cite{EK01.1} as a function of $\sigma_8$ for dark-matter halos at $z=0$ with three different masses, $M = 10^{14} M_\odot h^{-1}$ (black solid line), $M = 5 \times 10^{14} M_\odot h^{-1}$ (blue dashed line) and $M = 10^{15} M_\odot h^{-1}$ (red dot-dashed line).}
\label{fig:cmMzColl}
\end{figure}

This expectation is verified in the left panel of Figure~\ref{fig:cmMzColl}, where the halo concentration as obtained from the prescription by \cite{EK01.1} is shown as a function of halo mass at fixed redshift $z=0.3$, typical for strongly lensing clusters. Apart from the well-known decrease of concentration with halo mass, the figure shows that the concentration tends to increase with the normalisation. For example, the concentration of a cluster-sized dark-matter halo of $M=10^{15} M_\odot h^{-1}$ increases by $\sim 40\%$ when $\sigma_8$ is increased from $0.7$ to $0.9$. This fact alone may significantly affect arc statistics, because more compact cluster cores push the critical curves and caustics outwards, thus increasing their strong-lensing cross sections.

According to \cite{EK01.1}, the collapse redshift $z_\mathrm{c}$ of a halo of mass $M$ is implicitly given by
\begin{equation}\label{eqn:collapse}
  D_+(z_\mathrm{c}) \sigma(M_\mathrm{s}) \left[
    -\frac{d\ln\sigma(M_\mathrm{s})}{d\ln M} \right] = \frac{1}{C}\;,
\end{equation}
where $D_+$ is the linear growth factor for density fluctuations, normalised to unity at present, and $M_\mathrm{s}$ is the mass contained within the radius of maximum circular velocity for the NFW density profile, $2.17 r_\mathrm{s}$. The \emph{rms} density fluctuation $\sigma(M)$ is taken at the linear scale corresponding to $M$, and $C=28$ is a dimensionless constant calibrated against $N$-body simulations by \cite{EK01.1}.

\begin{figure}[t]
  \includegraphics[width=0.5\hsize]{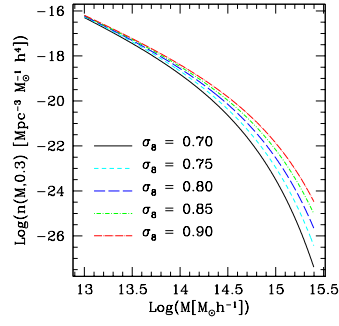}\hfill
  \includegraphics[width=0.5\hsize]{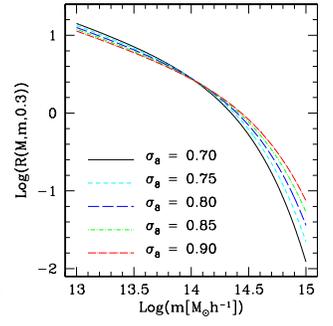}
\caption{\emph{Left panel}. \cite{PR74.1} mass function at fixed redshift $z=0.3$ as a function of mass. \emph{Right panel}. Merger rate between a cluster of mass $M = 10^{15} M_\odot h^{-1}$ and a substructure of mass $m$ indicated on the abscissa, at $z=0.3$. Both panels show results for different values of $\sigma_8$, as labelled. Note that here and in Figure~\ref{fig:z}, the merger rate is the probability for a dark-matter halo to merge with a substructure per unit logarithm of the merging mass and per unit logarithmic time.}
\label{fig:m}
\end{figure}

Evidently, higher $\sigma_8$ implies a lower growth factor at collapse redshift in equation~(\ref{eqn:collapse}), hence a higher collapse redshift. We show the collapse redshift for dark matter halos of different mass at redshift zero as a function of the $\sigma_8$ in the right panel of Figure~\ref{fig:cmMzColl}.

Alternative prescriptions for the computation of halo concentrations exist, but affect arc statistics only mildly. For instance, \cite{FE07.3} showed that the \cite{NA97.1} and \cite{BU01.1} recipes overestimate and underestimate respectively the cross section for giant arcs by a factor of $\sim 2$ with respect to the \cite{EK01.1} prescription. However the latter is probably the most general and physically best motivated. It turned out to reproduce halo concentrations in a variety of cosmologies, including those with dynamical dark energy \citep{DO04.2}.

\begin{figure}[t]
  \includegraphics[width=0.5\hsize]{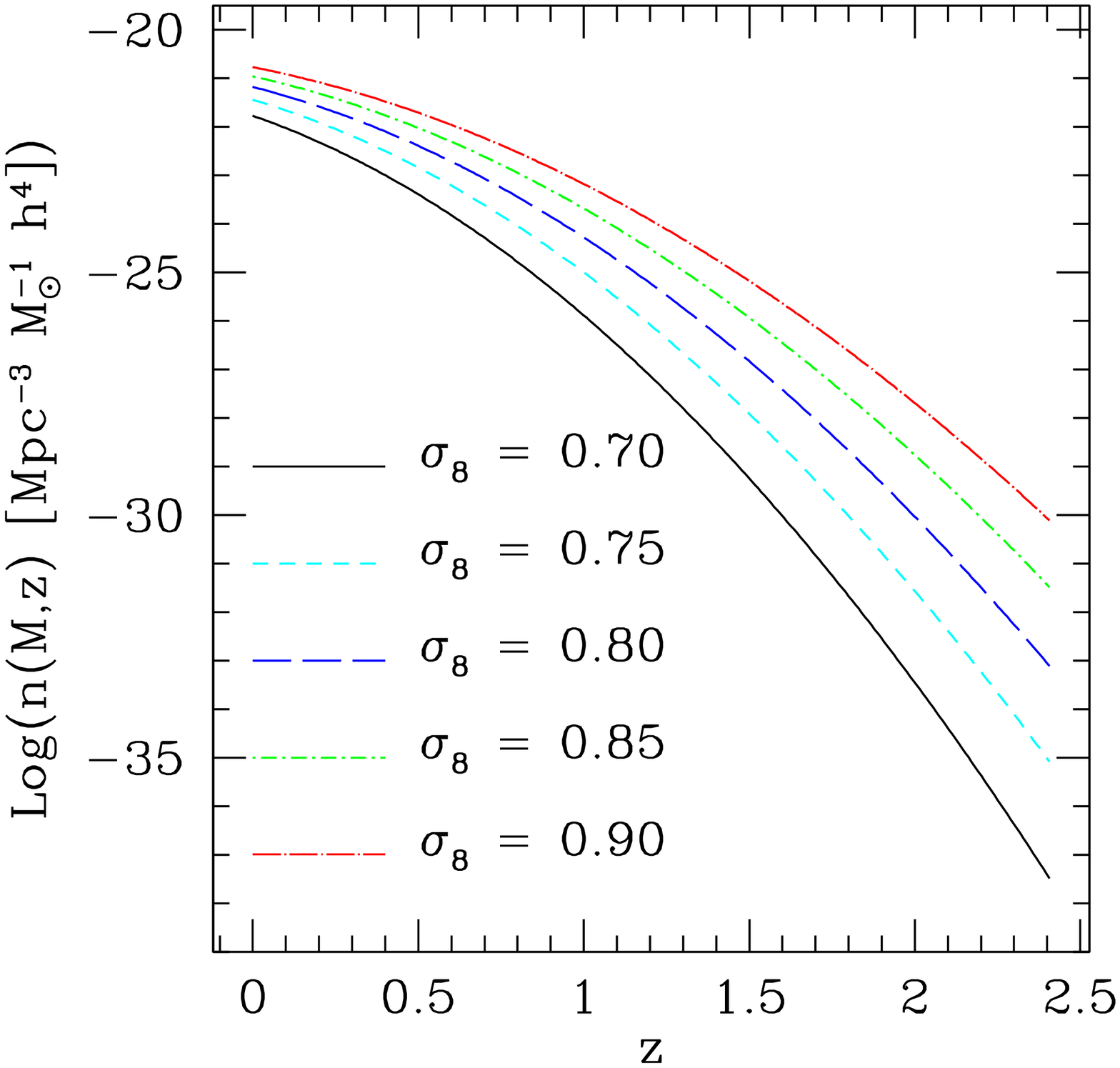}\hfill
  \includegraphics[width=0.5\hsize]{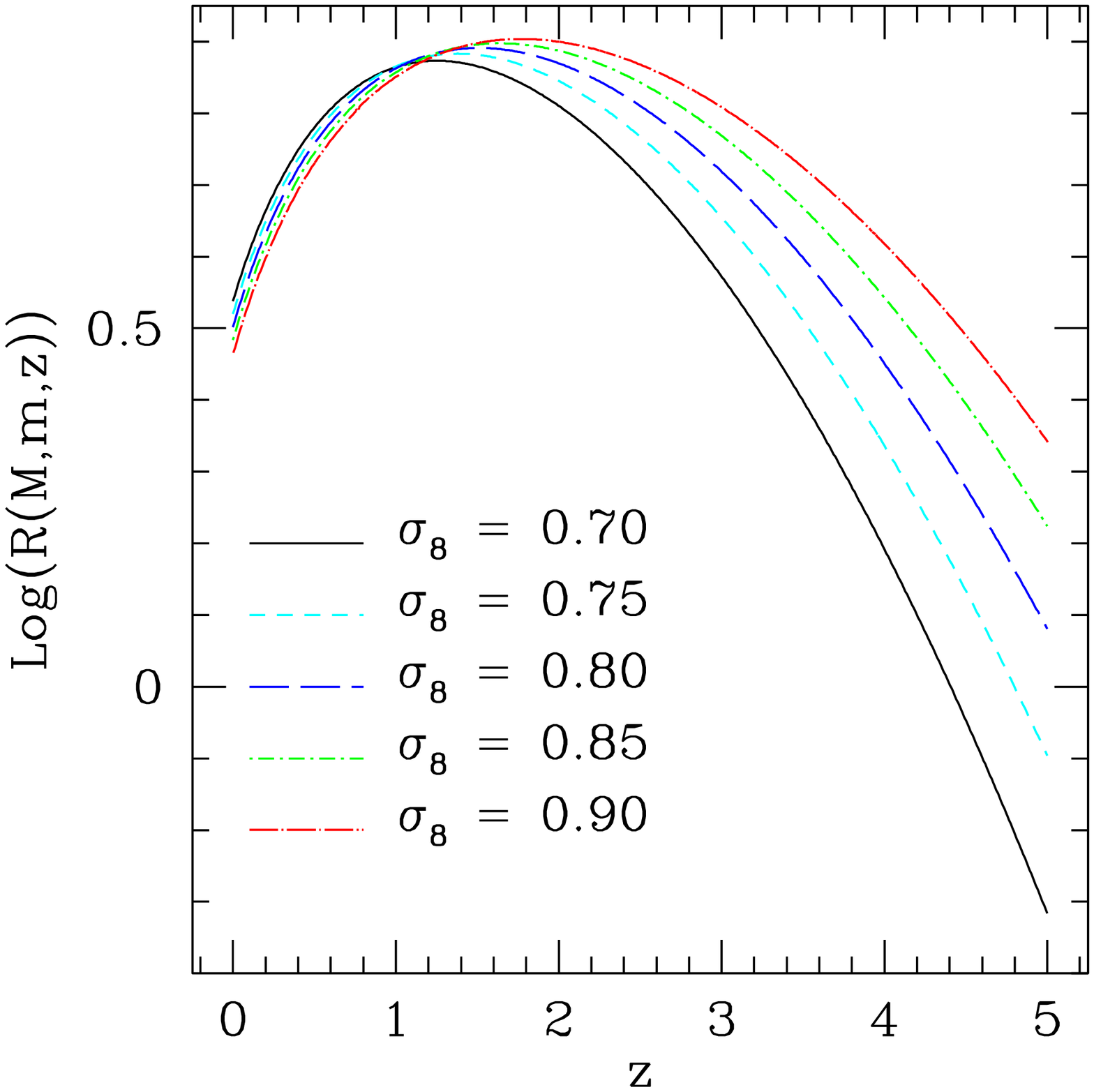}
\caption{\emph{Left panel}. Halo number density obtained from the \cite{PR74.1} mass function as a function of redshift at a fixed halo mass $M=7.5\times 10^{14} M_\odot h^{-1}$. \emph{Right panel}. Merger rate between a cluster of mass $M = 10^{15} M_\odot h^{-1}$ and a substructure of mass $m = 5 \times 10^{13} M_\odot h^{-1}$ as a function of redshift. Again, results are shown for five different values of $\sigma_8$.}
\label{fig:z}
\end{figure}

Next, we analyse the halo mass function and the merger rate. Figures~\ref{fig:m} and \ref{fig:z} show the \cite{PR74.1} mass function and the merger rate \citep{LA93.1,LA94.1} at fixed redshift as a function of mass, and at fixed mass as a function of redshift, respectively. We neglect the improvements \citep{JE01.1,SH02.1,WA06.2} of the mass function here because we only mean to illustrate the differences between different normalisations. The behaviour of the mass function is quite obvious. Higher normalisation gives rise to more structures at a given redshift. This is particularly evident at the high-mass end where the mass function depends exponentially on $\sigma_8^2$. At the massive cluster scale, the mass function can vary by orders of magnitude as $\sigma_8$ is varied.

The behaviour of the merger rate may be less obvious. It is larger for higher $\sigma_8$ if the mass of the main halo is large and the mass of the secondary halo is a considerable fraction of it. These are rare events because massive structures (and substructures) are rare. On the other hand, substructures much less massive than the main halo merge at a lower rate in highly normalised models. Because of the much higher abundance of low-mass halos, we expect mergers with low-mass substructures to be equally or even more frequent in models with lower $\sigma_8$, at least in the redshift interval relevant for our purposes. Since strong cluster lensing is highly sensitive to asymmetries and external perturbations, more merger events will further increase the strong-lensing cross sections.

Observed strong-lensing clusters frequently show substructures. While this could introduce a bias on the observational side, the study we shall refer to in this work is based on X-ray selected clusters \citep{LE94.1}. An investigation of the possible selection effects introduced by this fact, accounting for the boosting effect that cluster mergers have on both the lensing efficiency and the X-ray luminosity, is reported in \cite{FE07.2}. We are preparing a study of the strong-lensing properties of merging clusters in a large cosmological simulation.

The larger number-density of halos, their higher concentrations and the modified merger activity with increasing $\sigma_8$ will all combine to make the strong-lensing optical depth depend sensitively on the normalisation. It will be critically important for precise predictions to include the effect of mergers into the calculation. We shall now present our results, confirming these expectations.

\begin{figure*}[ht]
  \includegraphics[width=0.45\hsize]{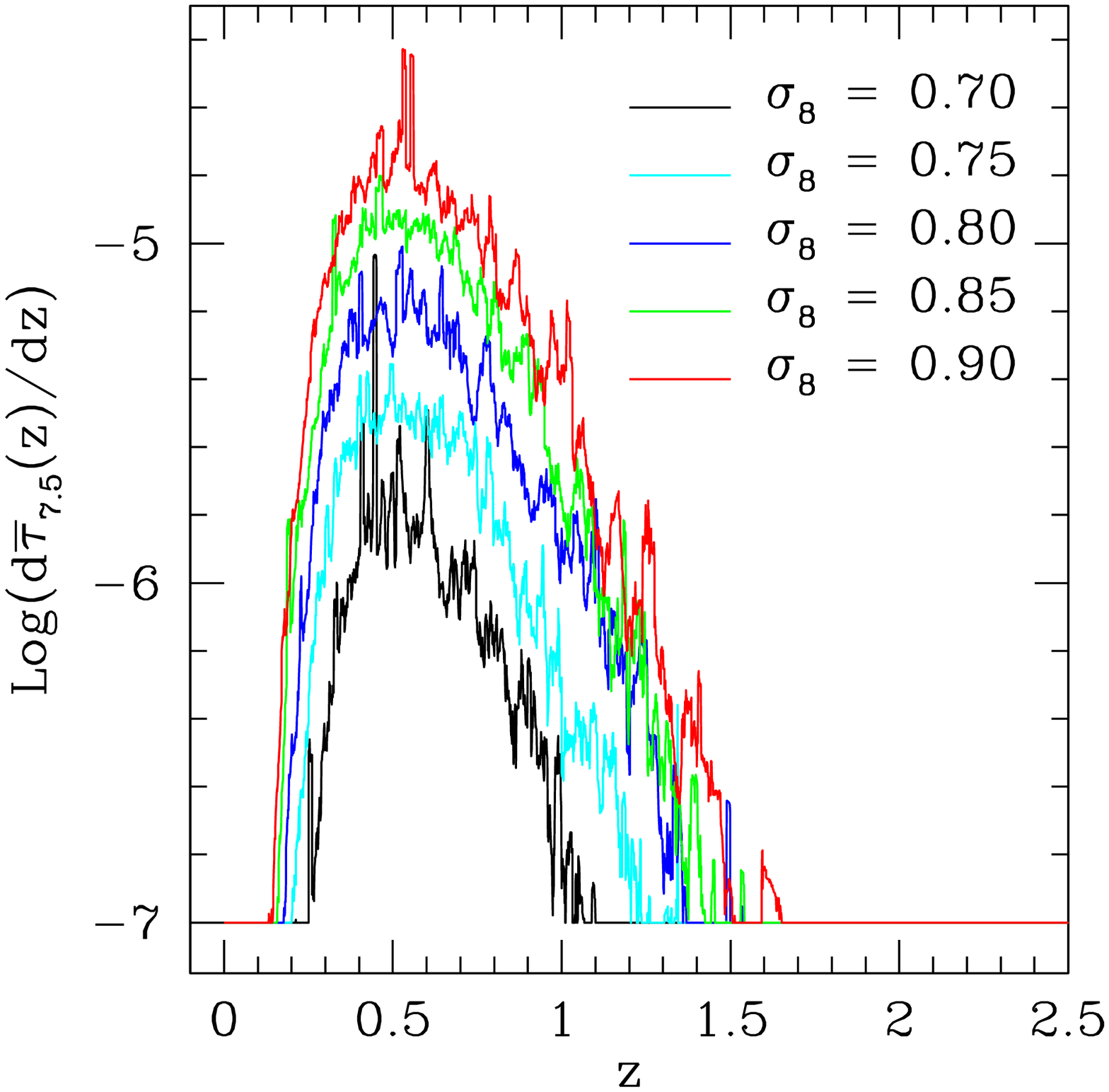}\hfill
  \includegraphics[width=0.45\hsize]{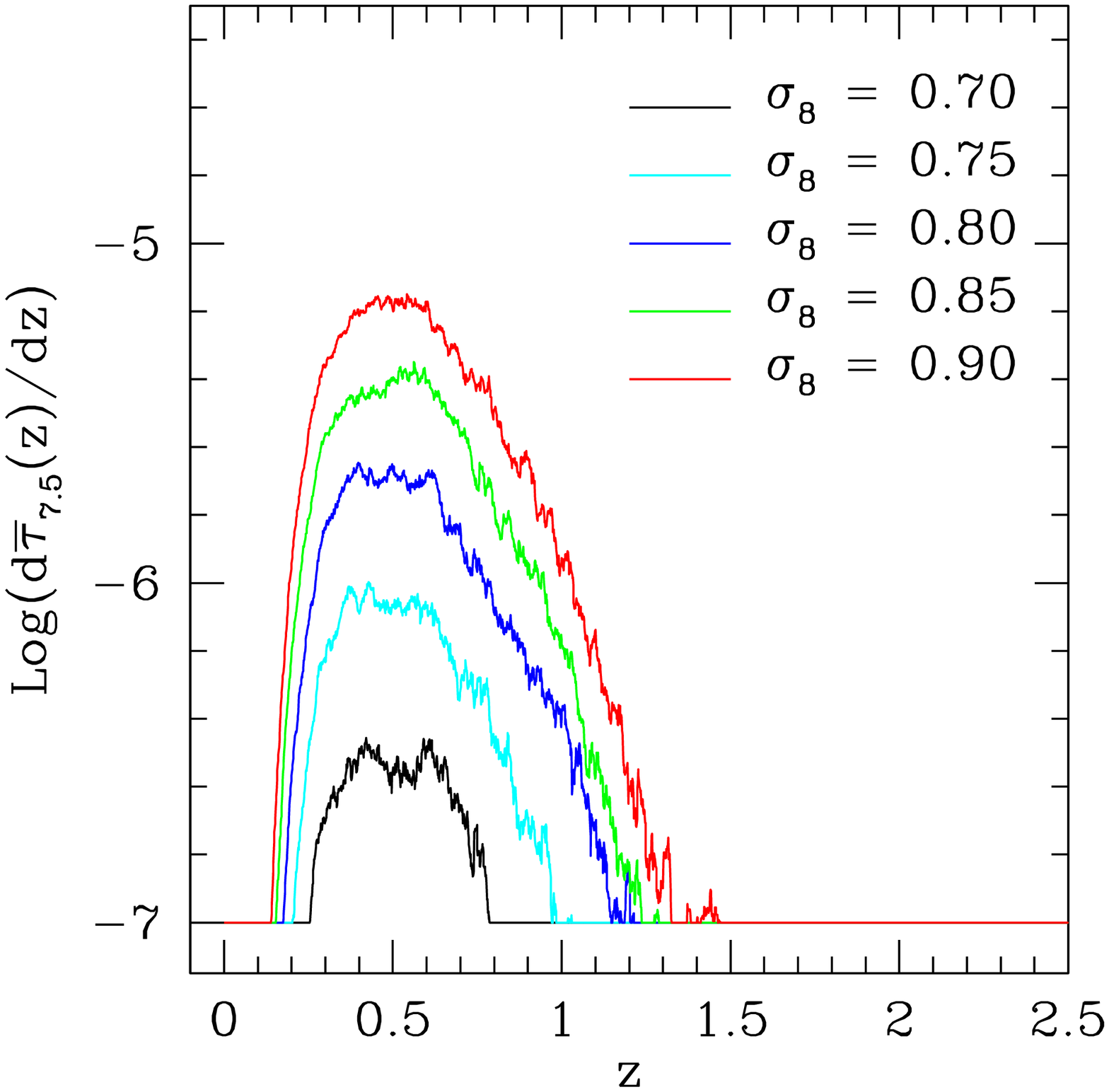}
\caption{The optical depth per unit redshift for arcs with length-to-width ratio $d\ge 7.5$ for five different values of $\sigma_8$, as labelled in the plots. The left and right panels show, on the same scale, results including and excluding the effect of cluster mergers, respectively. Predictions for $d\ge 10$ are extremely similar and therefore not shown here.}
\label{fig:tau}
\end{figure*}

\section{Cluster population}\label{sct:clu}

To produce realistic models of the cluster population without time-consuming numerical simulations, we adopt the extended Press-Schechter formalism \citep{PR74.1,BO91.1,LA93.1} to construct merger and formation histories for a set of $\mathcal{N} = 1,000$ dark-matter halos for each model universe. Examples for the application of this procedure are given in \cite{RA02.1} and \cite{FE07.1}, among others.

To cover the mass range relevant for strong cluster lensing, we draw the dark-matter halos uniformely from the mass interval between $10^{14} M_\odot h^{-1}$ and $2.5 \times 10^{15} M_\odot h^{-1}$ at redshift $z=0$. Each object is then evolved backwards in time in suitably chosen discrete time steps. Lensing by each dark-matter halo at each redshift step is modelled using a NFW density profile whose lensing potential is elliptically deformed with an ellipticity of $\epsilon=0.3$. This value was shown to give the best agreement with deflection angle maps of realistic simulated clusters \citep{ME03.1}.

The transient boost of the strong-lensing efficiency due to cluster mergers is taken into account by modelling the merging substructures also as elliptical NFW lenses. Each time a main cluster halo undergoes a merger with a substructure having more than $5\%$ of its mass, the encounter is modelled assuming that the two halos approach at a constant velocity starting from a distance equal to the sum of their virial radii, and assuming that the merger proceeds at the gravitational free-fall time.

Given the deflection-angle maps for each model cluster at each redshift, we compute the strong-lensing cross section for arcs with length-to-width ratio $d\ge7.5$ and $d\ge10$. Source redshifts were drawn randomly from the distribution
\begin{equation}\label{eqn:zs}
  p(z_\mathrm{s})=\frac{\beta}{z_0^3\Gamma(3/\beta)}\,
  z_\mathrm{s}^2\exp\left[-\left(
    \frac{z_\mathrm{s}}{z_0}
  \right)^\beta\right]\;,
\end{equation}
\citep{SM95.1}, with $z_0=1$ and $\beta=3/2$. The distribution peaks at $z\sim 1.2$, implying that objects at $z \simeq 0.3 - 0.5$ are the most efficient lenses. Model clusters were evolved backwards in time up to the source redshift.

Photometric redshifts measured for recent wide-area surveys \citep{IL06.1,SE06.1} favour somewhat different values for the parameters of equation (\ref{eqn:zs}) or even a different functional form for the redshift distribution, implying a peak at $z \lesssim 0.7$. While such distributions would be significantly less efficient in producing large amounts of long and thin arcs, it is likely that these wide surveys are not deep enough to capture the complete source population relevant for the production of strong-lensing features. This is suggested by the high measured redshifts of many giant arcs in galaxy clusters (see e.g. \citealt{EL07.1} for a recent example).

We compute the lensing cross sections by means of the semi-analytic prescription by \cite{FE06.1} instead of costly ray-tracing simulations. The method consists of integrating the inverse magnification along critical curves over an area determined by the locally linearised lens mapping.
In particular, the area is defined by the condition $|\lambda_\mathrm{r}/\lambda_\mathrm{t}| \ge d$, where $\lambda_\mathrm{r}$ and $\lambda_\mathrm{t}$ are the (radial and tangential) eigenvalues of the lensing Jacobian matrix. Finite source size is accounted for by convolving the lens properties with functions standing for circular background galaxies with $0.5"$ radius, and source ellipticity is included in the computation by using the formalism developed by \cite{KE01.2}. Further detail is given in \cite{FE06.1}.

Having computed the strong-lensing cross section of each individual cluster at each redshift step, we can compute the differential optical depth per unit redshift $\d\bar{\tau}_\mathrm{d}/\d z$. As mentioned, we use the common thresholds $d=7.5$ and $d=10$ for the length-to-width ratio, but show results for only one of them if the predictions have similar behavior. Since the cluster population is represented by a discrete set of objects,
\begin{equation}\label{eqn:tau}
  \frac{\d\bar{\tau}_\mathrm{\d}(z)}{\d z} = \sum_{i=1}^{\mathcal{N}-1}
    \frac{\sigma_\mathrm{d}(M_i,z,z_{\mathrm{s},i})}{4 \pi D_{\mathrm{s},i}^2}
    \int_{M_i}^{M_{i+1}} n(M,z)\d M\;,
\end{equation}
where $D_{\mathrm{s},i}$ is the angular-diameter distance to the source sphere of the $i$-th cluster, and $n(M,z)$ is the number of structures with mass within $M$ and $M+\d M$, and redshift within $z$ and $z+\d z$. The masses $M_i$ are assumed to be in ascending order, $M_{i+1} \ge M_i$ for all $i$. Since source redshifts are assigned randomly to each cluster in our synthetic sample, the weighting with the redshift distribution is implicitely included in the calculation.

The integral over the lens redshift of equation~(\ref{eqn:tau}) is the average optical depth. After multiplication with the total number of available sources in the sky, it yields the total number of gravitational arcs with length-to-width ratio $\ge d$,
\begin{equation}\label{eqn:num}
  N_\mathrm{d} = n_\mathrm{s} \bar{\tau}_\mathrm{d} \equiv n_\mathrm{s}
  \int_0^{+\infty} \frac{\d\bar{\tau}_\mathrm{d}(z)}{\d z} \d z\;.
\end{equation}
It should be noted that the integral on the r.h.s. of equation~(\ref{eqn:num}) cannot extend to infinity for our discrete cluster sample, but only to a finite value $z_\mathrm{max}$ where the probability of finding sources according to the adopted redshift distribution equation~(\ref{eqn:zs}) is negligible. We adopted $z_\mathrm{max} = 7.5$ (see also the discussion in \citealt{FE07.1}).

\begin{figure}[t]
  \includegraphics[width=\hsize]{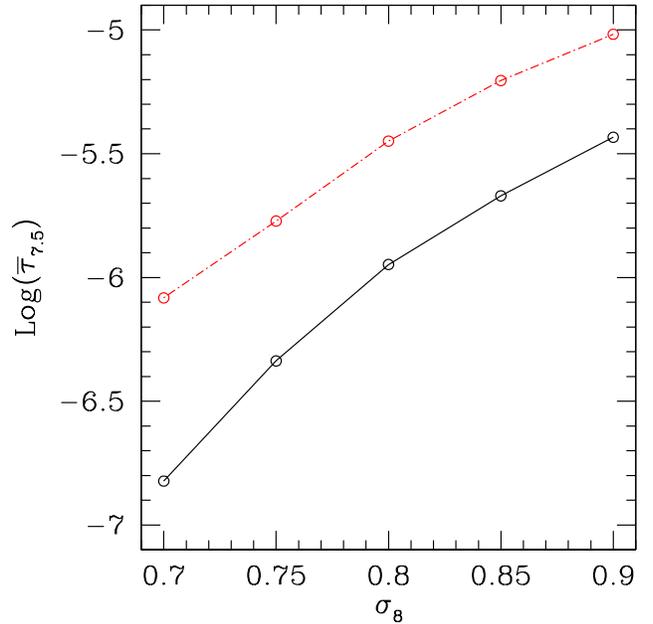}
\caption{The total average optical depth for arcs with length-to-width ratio $d\ge7.5$ as a function of $\sigma_8$. The red dot-dashed line includes mergers, while the black solid line ignores them.}
\label{fig:opt}
\end{figure}

\begin{figure}[ht]
  \includegraphics[width=\hsize]{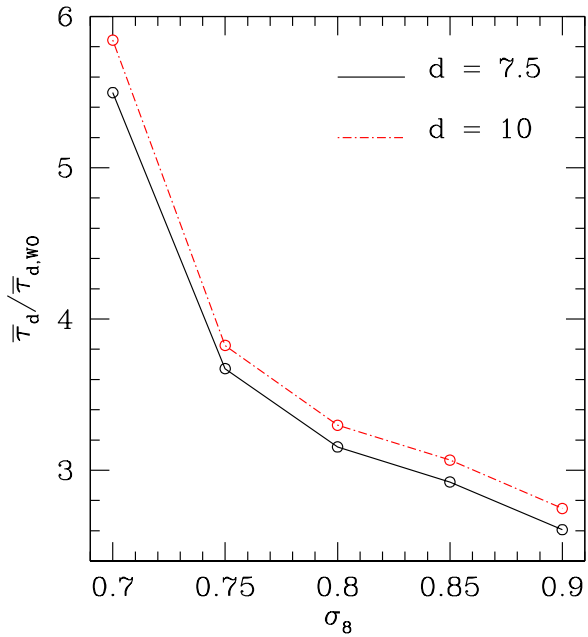}
\caption{The ratio of the total average optical depth for arcs with length to width ratio larger than or equal to $d = 7.5$ (black solid line) and $d=10$ (red dot-dashed line) to that obtained ignoring the boosting effect of cluster mergers, as a function of $\sigma_8$.}
\label{fig:ratio}
\end{figure}

\section{Results}

\subsection{Optical Depth}

Figure~\ref{fig:tau} illustrates the strong-lensing effects of our synthetic cluster population. It shows the optical depth per unit redshift given by equation~(\ref{eqn:tau}), computed for the five different normalisations $\sigma_8$ analysed here. We also show the change in optical depths caused by cluster mergers. Lines are drawn only for one of our choices of the length-to-width threshold for gravitational arcs, namely $d=7.5$. Differential optical depths for $d=10$ are slightly smaller, but the qualitative behavior is unchanged.

As anticipated in Sect.~\ref{sct:exp}, the optical depth per unit redshift increases substantially with $\sigma_8$ because more halos with more concentrated mass distributions exist as $\sigma_8$ increases. This difference is particularly strong at high redshifts, $z \gtrsim 1$, where the lensing efficiency is still significant for $\sigma_8 = 0.9$ while being negligible for $\sigma_8 = 0.7$. In this sense, a higher normalisation is degenerate with the introduction of an early-dark energy component (see the discussion in \citealt{FE07.1,FE07.2}). 

Figure~\ref{fig:opt} shows the total optical depth according to equation~(\ref{eqn:num}), i.e.~the integral over each of the curves shown in Figure~\ref{fig:tau}, for five values of $\sigma_8$, either including or neglecting cluster mergers. There is more than one order of magnitude difference between the model with lowest and the highest normalisations, $\sigma_8 = 0.7$ and $\sigma_8 = 0.9$, independent of whether cluster mergers are included. This high sensitivity of the expected number of arcs on $\sigma_8$ has important consequences for the arc statistics problem, as will be discussed in detail in Sect.~\ref{sct:con}.

\begin{figure*}[ht]
  \includegraphics[width=0.45\hsize]{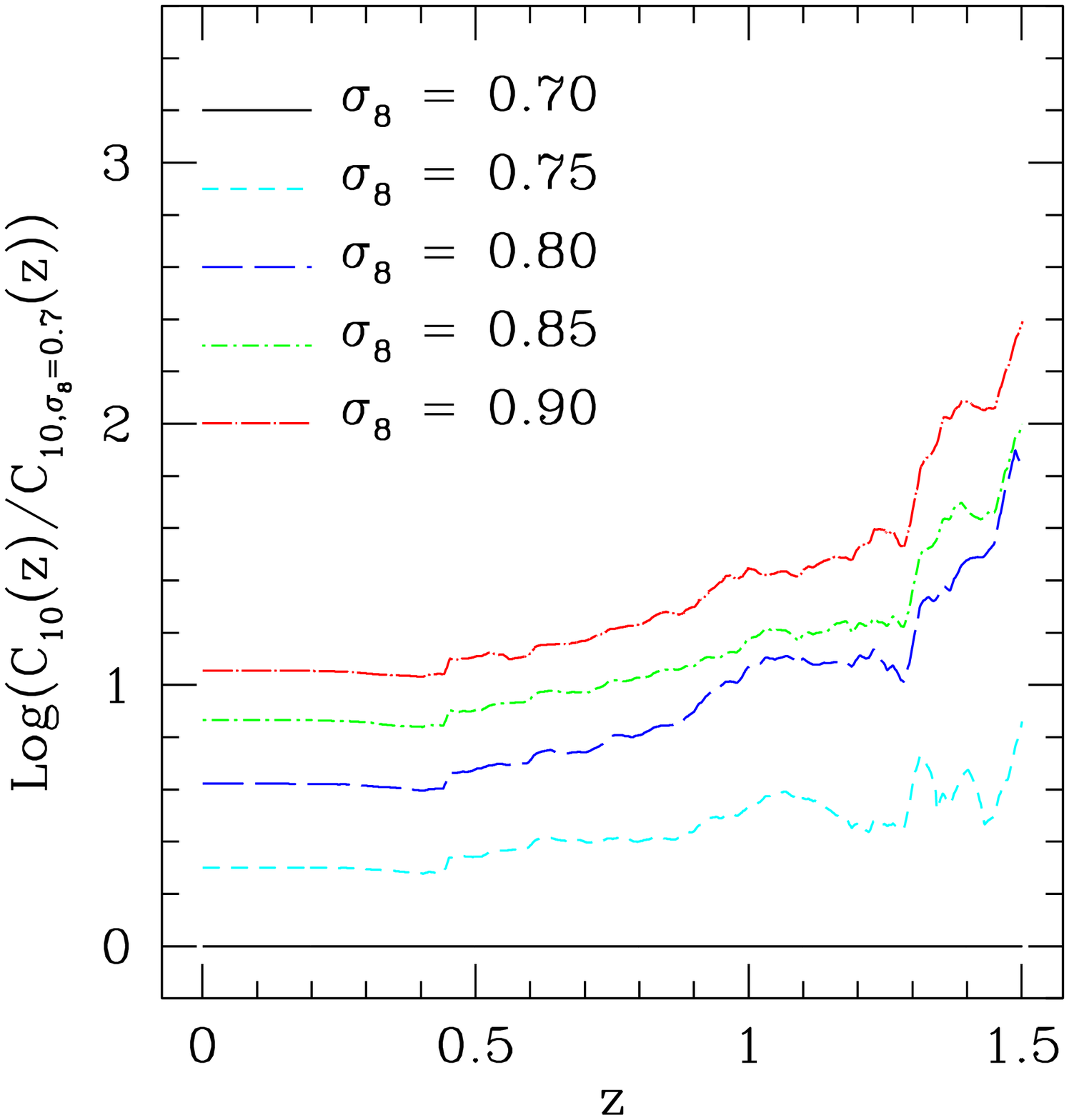}\hfill
  \includegraphics[width=0.45\hsize]{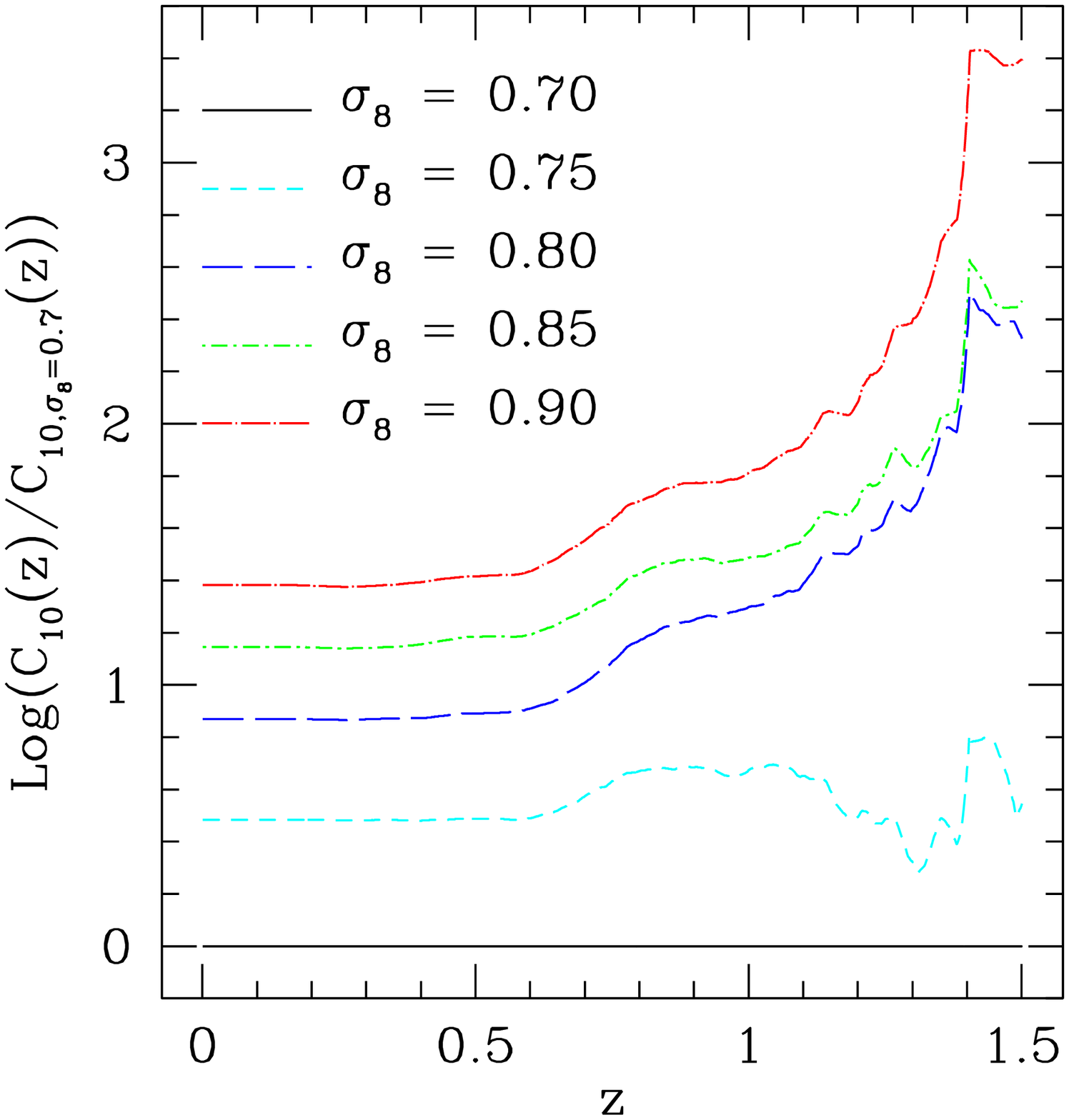}
\caption{The cumulative optical depth for different $\sigma_8$ is shown relative to its value for $\sigma_8=0.7$. The left and right panels show results including and neglecting cluster mergers, respectively. The results are quite independent of the length-to-width threshold; $d\ge10$ was chosen here.}
\label{fig:cumulative}
\end{figure*}

Figure~\ref{fig:opt} also shows that mergers are more important when $\sigma_8$ is low. This is emphasised in Figure~\ref{fig:ratio}, which shows the ratio between the total optical depth obtained including and neglecting cluster mergers. Evidently, dynamical activity enhances the optical depth by a factor $\gtrsim5$ if $\sigma_8$ is low, and by a factor $\gtrsim3$ when $\sigma_8$ is high. This confirms results obtained earlier by \cite{FE07.1}. We also note that the effect of mergers is high for more extreme arcs, $d\ge10$. This is because the individual cross sections are smaller in this case and therefore relatively more sensitive to perturbations.

It is also highly interesting to see how the strong-lensing efficiency of clusters at high redshift changes with $\sigma_8$. This is particularly relevant in view of the apparent unexpectedly high incidence of gravitational arcs in distant galaxy clusters \citep{GL03.1,ZA03.1}. We quantify this by means of the cumulative optical depth
\begin{equation}\label{eqn:cumulative}
  C_\mathrm{d}(z)\equiv\int_z^{+\infty} \frac{\d\bar{\tau}_\mathrm{d}(z')}{\d z'}\d z'
\end{equation}
contributed by clusters above redshift $z$ (the upper integration limit is set to $z_\mathrm{max} = 7.5$ here as well). Figure~\ref{fig:cumulative} shows $C_\mathrm{d}(z)$ for different normalisations $\sigma_8$ \textit{relative} to $C_\mathrm{d}(z)$ for $\sigma_8=0.7$, including and neglecting mergers. This ratio is virtually independent of the length-to-width threshold, so we show it for $d\ge10$ only.

The relative contribution to the optical depth increases towards high redshifts for all models with $\sigma_8>0.7$. The increase is higher when mergers are ignored ($\sim3$ instead of $\sim2$ orders of magnitude) because the relative effect of mergers decreases as unperturbed clusters by themselves become stronger lenses. The main conclusion is that high $\sigma_8$ makes gravitational arcs at high redshift extremely more likely. Conversely, this means that arcs in distant clusters are a massive problem for models with low $\sigma_8$.

\subsection{Number of Arcs}

We now need to transform our predictions for the optical depth, including mergers and a realistic source redshift distribution, into predictions for the numbers of observable arcs on the sky. To do this, it suffices in principle to multiply the average optical depth $\bar{\tau}_\mathrm{d}$  with the total number of sources in the sky, according to equation~(\ref{eqn:num}).

However, it is necessary to include a luminosity function for the faint blue background galaxy population into the calculation and to account for the magnification effect due to gravitational lensing. The latter has a twofold impact on the observed source counts. First, faint sources are magnified above the flux threshold for detection, thus increasing the number of sources visible per unit solid angle. Second, the sky is locally stretched, thus reducing the number density of the source galaxies. If the original flux distribution function of the sources is a power law with logarithmic slope $-1$, then the two effects of flux magnification and number dilution cancel, leaving the number of observed sources per unit solid angle unchanged \citep{BA01.1}.

\begin{figure}[t]
  \includegraphics[width=\hsize]{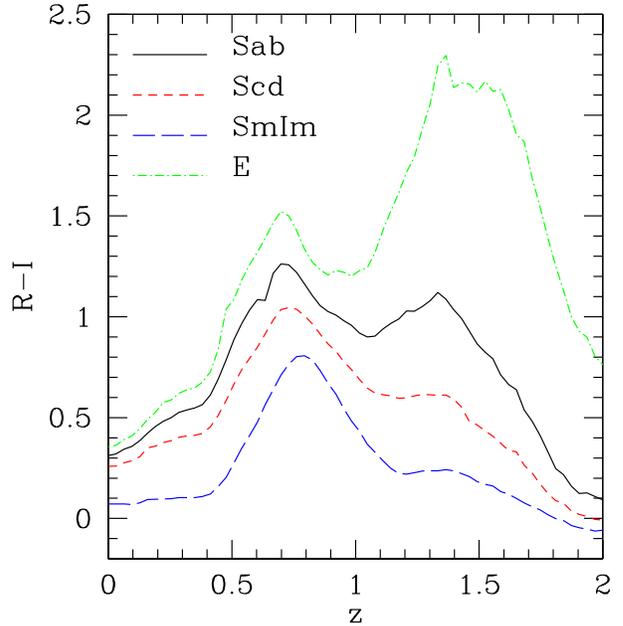}
\caption{The $\mathrm{R}-\mathrm{I}$ colour as a function of redshift, computed from the spectra of three different morphological types of spirals and for elliptical galaxies, as labelled in the plot.}
\label{fig:colours}
\end{figure}

We model the number counts of faint background galaxies as a function of the observed apparent magnitude to match the measurements of \cite{CA00.1} in the Hubble Deep Field. There, only the I-band magnitudes are used, and the number counts are fitted using the relation
\begin{equation}\label{eqn:mag}
  n_0(m_\mathrm{I}) = \frac{n_{0,*}}{\sqrt{10^{2a(m_{\mathrm{I},1}-m_\mathrm{I})} + 
  10^{2b(m_{\mathrm{I},1}-m_\mathrm{I})}}}\;,
\end{equation}
with the best-fit parameters $a=0.30$, $b=0.56$, $m_{\mathrm{I},1}=20$ and $n_{0,*} = 3 \times 10^{3}$ per square degree and per unit magnitude. Since these number counts are given in the I-band only, we convert them to the R-band by using the approximate relation $\mathrm{R} \simeq \mathrm{I}+ 1$. In Figure~\ref{fig:colours} the colour $\mathrm{R}-\mathrm{I}$ is shown as a function of redshift for different morphological types of galaxy. Since galaxies imaged as long and thin arcs are usually blue spirals (see \citealt{TY88.1,EL97.1}), $\mathrm{R}-\mathrm{I}\sim 1$ approximately holds over the redshift range $z \lesssim 1.7$. However, also adopting the relation $\mathrm{R} - \mathrm{I} \sim 0.5$, which holds better at higher redshift, the magnified number counts (see below) change by less than $50\%$, which is irrelevant given the uncertainties in the original galaxy counts \citep{CA00.1}. We thus conclude that the approximation $\mathrm{R} \simeq \mathrm{I}+ 1$ is admittedly rough, but sufficient for our purposes.

\begin{figure}[t]
  \includegraphics[width=0.5\hsize]{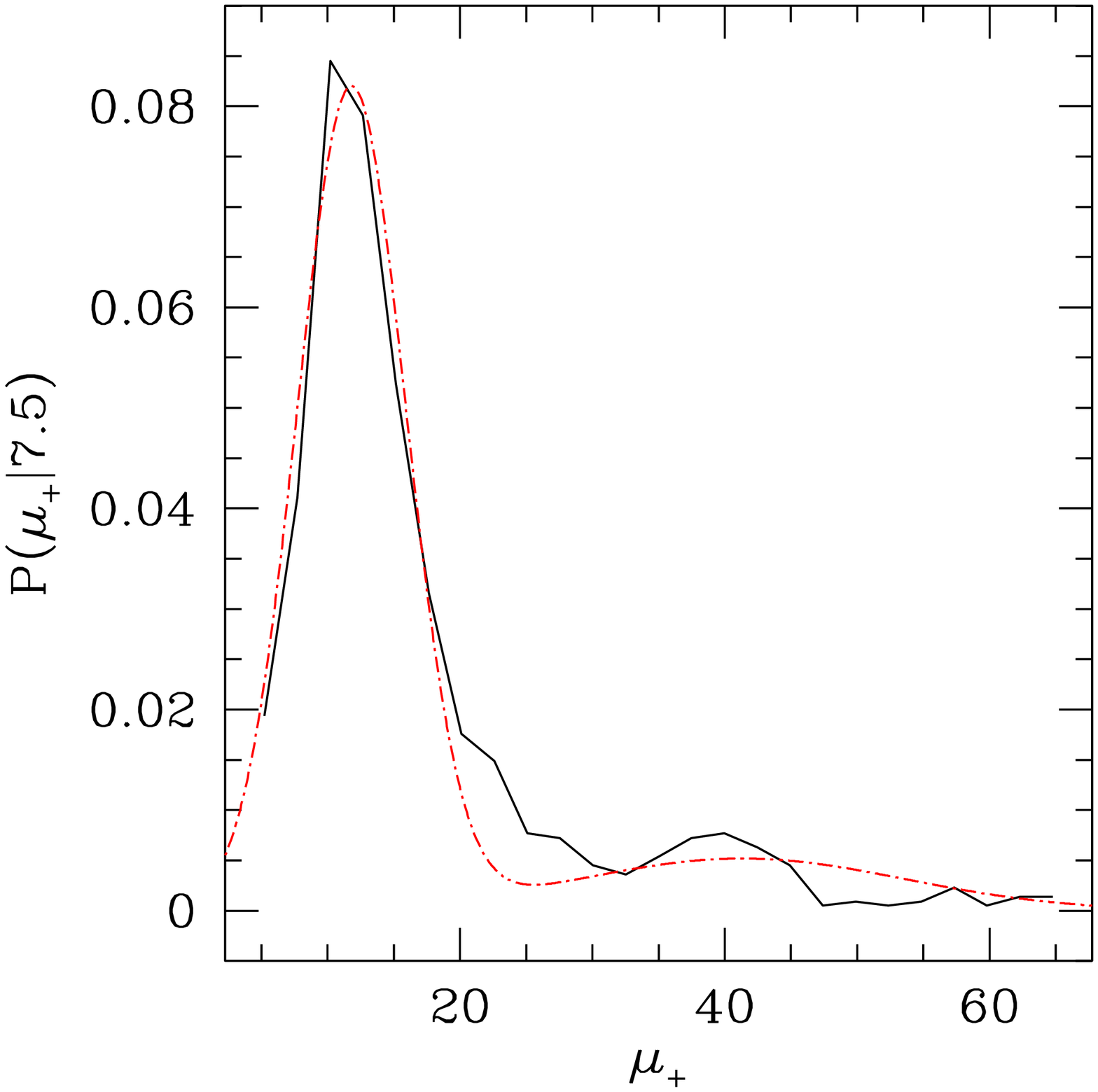}\hfill
  \includegraphics[width=0.5\hsize]{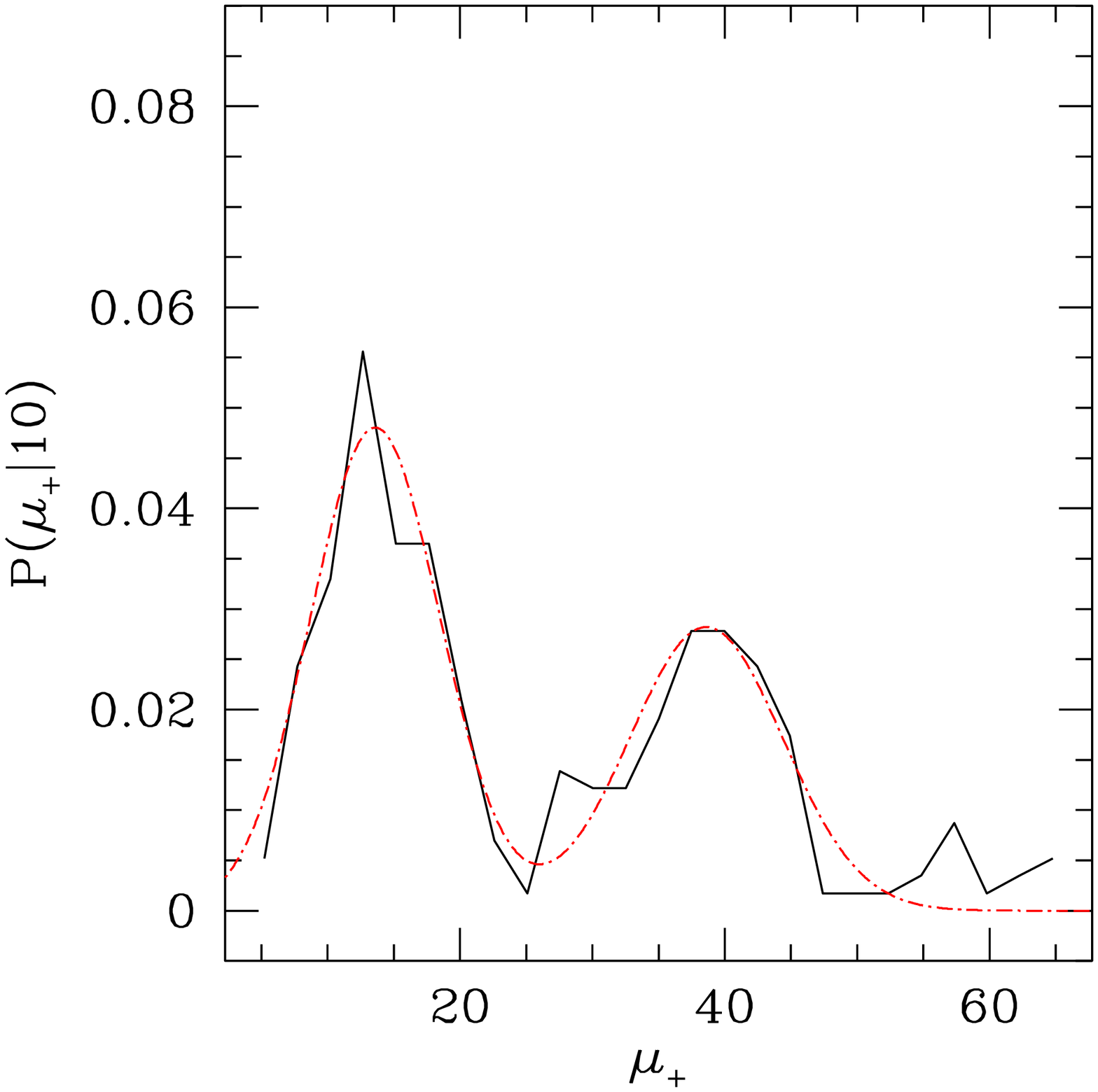}
\caption{Conditional probability distribution for the magnification of images given a threshold for the length-to-width ratio of $d=7.5$ (left panel) and $d=10$ (right panel). The black solid lines show the result of ray-tracing simulations, while the red dot-dashed curves are the best-fitting two-component Gaussians whose parameters are summarised in Tab.~\ref{tab:2gaus}.}
\label{fig:conditional}
\end{figure}

Figure~\ref{fig:conditional} shows the conditional probability distribution for the magnification of background sources given the length-to-width threshold for the imaged arcs. The results shown are obtained from fully numerical ray-tracing simulations, to which we fit two-component Gaussians for both thresholds $d=7.5$ and $d=10$. The two-component Gaussian is
\begin{eqnarray}
P(\mu_+|d) &=& \frac{A}{\sqrt{2\pi}\sigma_1} \exp \left[
  -\frac{(\mu_+-\mu_{+,1})^2}{2\sigma_1^2} \right] +
\nonumber\\
&+&\frac{1-A}{\sqrt{2\pi}\sigma_2} \exp \left[
  -\frac{(\mu_+-\mu_{+,2})^2}{2\sigma_2^2} \right],
\end{eqnarray} 
with $\mu_+ \equiv |\mu|$. The best-fit parameters $A$, $\sigma_i$ and $\mu_{+,i}$ are summarised in Tab.~\ref{tab:2gaus} for both $d=7.5$ and $d=10$. Obviously, the magnification is not a good estimator for the length-to width ratio of an image, especially when high length-to-width thresholds are used.

\begin{table}[b]
\caption{Parameters for the two-component Gaussian fit to the conditional probability distribution for magnification given a threshold $d$ for the length-to-width ratio of simulated gravitational arcs.}
  \label{tab:2gaus}
  \begin{center}
    \begin{tabular}{|l|l|l|}
      \hline
      Parameter & $d=7.5$ & $d=10$\\
      \hline
      \hline
      $A$ & $0.84$ & $0.59$ \\
      $\sigma_1$ & $4.1$ & $4.9$ \\
      $\sigma_2$ & $12.3$ & $5.8$ \\
      $\mu_{+,1}$ & $11.8$ & $13.6$ \\
      $\mu_{+,2}$ & $41.3$ & $38.6$ \\
      \hline
    \end{tabular}
  \end{center}
\end{table}

\begin{figure*}[ht]
  \includegraphics[width=0.45\hsize]{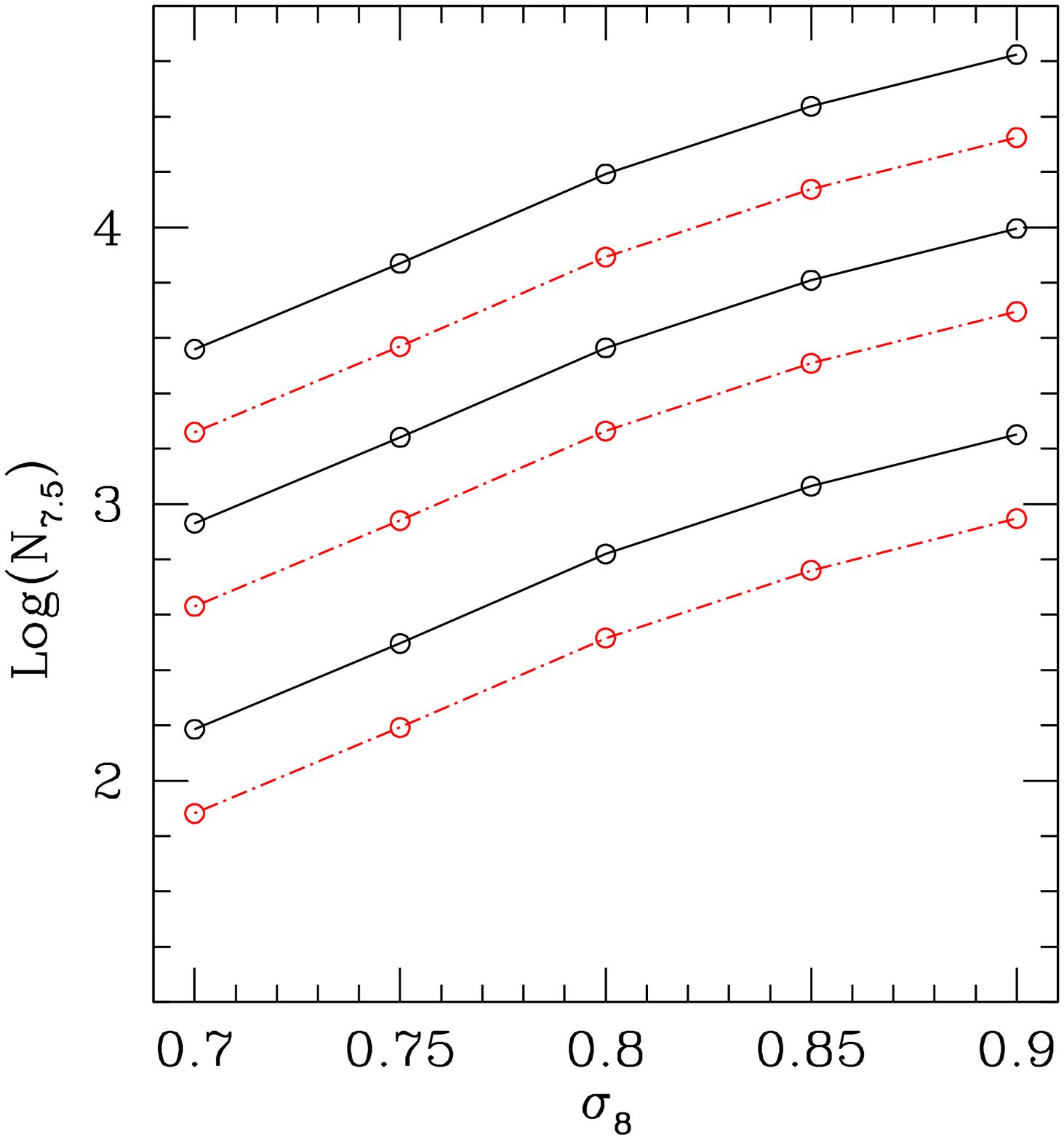}\hfill
  \includegraphics[width=0.45\hsize]{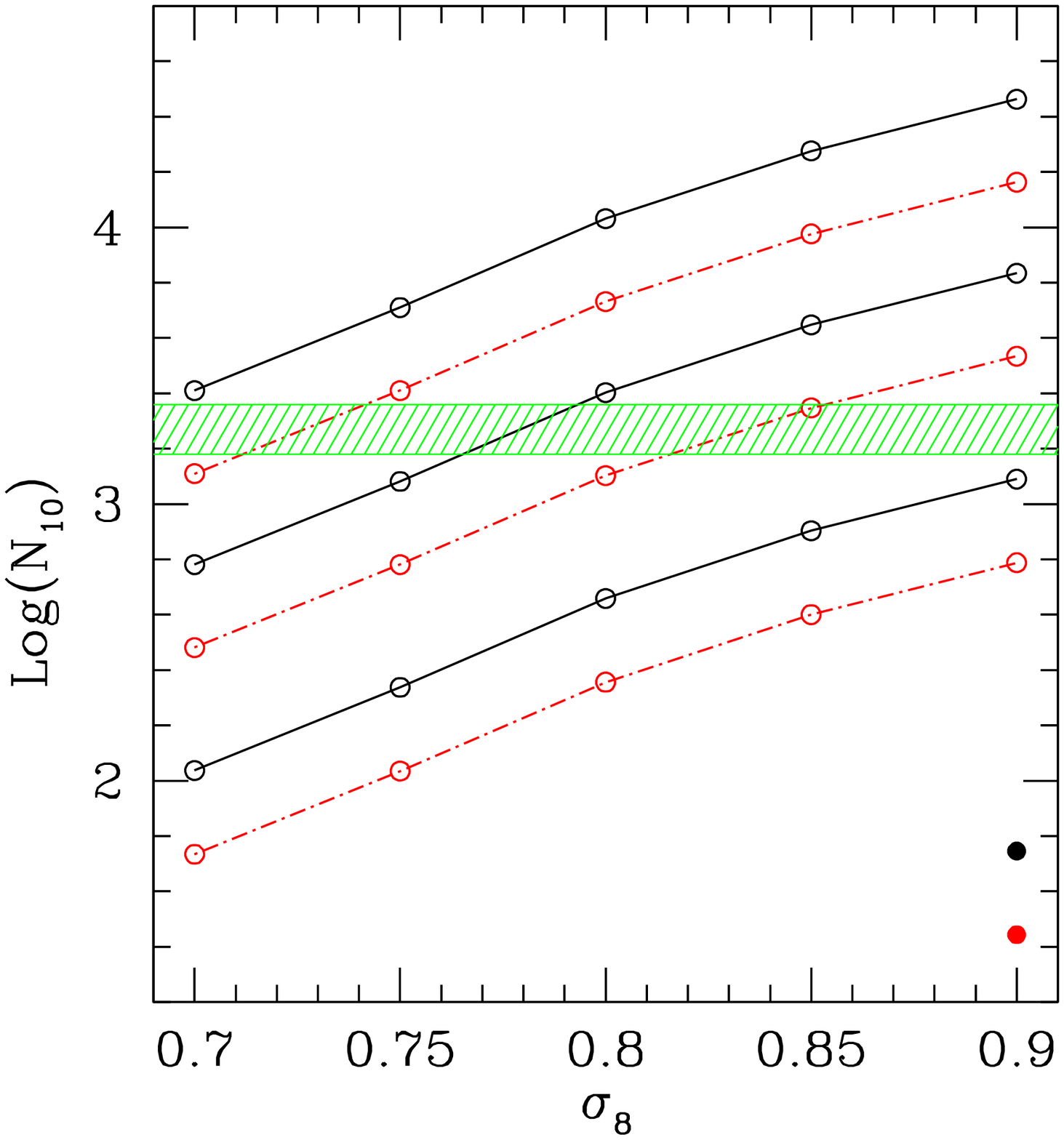}
\caption{The number of arcs with length-to-width ratio $d\ge7.5$ (left panel) and $d\ge10$ (right panel) predicted to be observable in a $\Lambda$CDM model normalised by $\sigma_8$. Black solid lines refer to I-band, red dot-dashed lines to R-band magnitudes. Results for three different limiting magnitudes are shown: 21.5 (bottom pair of curves), 24 (middle pair) and 26 (top pair). The right panel also shows the observed number of giant arcs according to \cite{LE94.1,GI94.1} as a green shaded area, to be compared with the lowest red curve. Moreover, the filled black and red points represent the original result of \cite{BA98.2} (in the I and R bands, respectively), rescaled to the source number counts used in this work.}
\label{fig:nArcs}
\end{figure*}

The original number counts read off \cite{CA00.1} are thus convolved with the conditional probability distribution for arcs with thresholds $d=7.5$ and $d=10$ to obtain the number counts after the magnification bias. Let $n_0(F)$ be the original flux distribution function for the sources, that is the number of sources per unit solid angle contained in the unit flux interval around $F$. It corresponds to the magnitude distribution of equation~(\ref{eqn:mag}) with the magnitude replaced by the flux. Then the magnified distribution is
\begin{equation}
  n(F) = \int_0^{+\infty} n_0\left( \frac{F}{\mu_+} \right)
  \frac{P(\mu_+|d)}{\mu_+^2} d\mu_+\;.
\end{equation}
The magnified number counts can then simply be multiplied with the average optical depth to find the total number of arcs. Figure~\ref{fig:nArcs} shows the total number of arcs with length-to-width ratio $d\ge7.5$ and $d\ge10$ predicted to be observable on the whole sky as a function of $\sigma_8$. Results are shown for three different limiting magnitudes in both the I and R bands. Arc surveys in X-ray selected cluster samples focused in the past on giant arcs, i.e. arcs with length-to-width ratio $d\ge10$ and R-band magnitudes less than $\mathrm{R}_\mathrm{lim} =21.5$, finding $\sim 10^3$ such arcs extrapolated to the whole sky \citep{LE94.1,BA98.2}.

The numbers given in Figure~\ref{fig:nArcs} clearly show how the prediction falls short of the observation for all values of $\sigma_8$ considered here, including $\sigma_8=0.9$. In the latter case, however, the difference is only a factor of $\sim 2$ and can possibly be accommodated including minor contributions due to cluster galaxies \citep{ME00.1,FL00.1} or a slightly different parametrisation for the source-redshift distribution. Much progress has been made since the first prediction by \cite{BA98.2}, which is also included in the Figure. However, if $\sigma_8$ is close to the value inferred from the WMAP-3 data, all this progress could not alleviate the arc statistics problem: with $\sigma_8 \sim 0.75$, the predicted number of arcs still falls about one order of magnitude below the observed number. Should $\sigma_8\sim0.8$ persist, there is still a factor $\sim6.5$ between the prediction and the observation.

For completeness, we repeat the calculation of the optical depth for arcs with length-to-width ratio $d\ge10$ including the scatter in the relation between the mass and the concentration of each dark-matter halo in the population. The procedure used is the same as described in \cite{FE07.3}, to which we refer for details. We just recall that the distribution of concentrations around the nominal value for a given halo mass is well fitted by a log-normal distribution with a \emph{rms} of $\sigma_c \sim 0.2$ \citep{JI00.1}. It was shown in \cite{FE07.3} that the scatter in the mass-concentration relation may increase the total, average optical depth by up to $40-50\%$. It was also shown there that additional effects like the dependence of the concentration on the triaxiality of the dark-matter halos and the distribution of projected ellipticities due to the triaxiality itself do not introduce any additional bias into these results.

\begin{figure}[t]
  \includegraphics[width=\hsize]{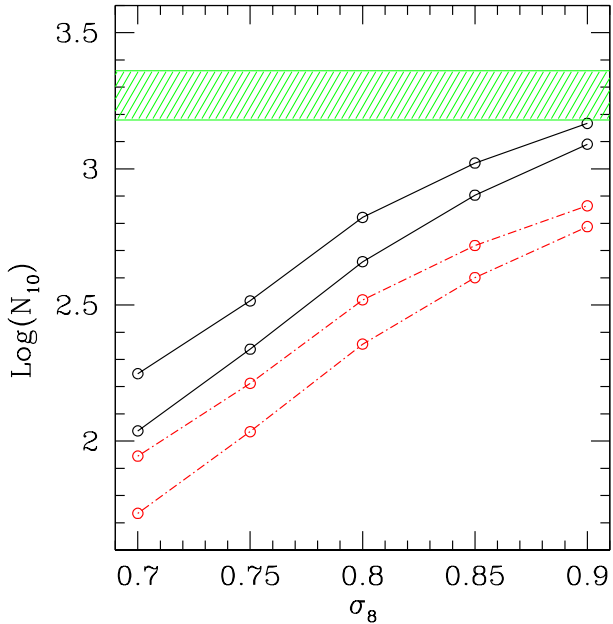}
\caption{The number of arcs with length-to-width ratio $d\ge10$ and magnitude $\le21.5$ predicted to be observable in a $\Lambda$CDM universe as a function of $\sigma_8$, taking cluster mergers into account. Black solid curves refer to I-band, red dot-dashed curves to R-band magnitudes. The higher curve of each pair contains the scatter in the relation between halo masses and concentrations, while the lower ones are reproduced from Figure~\ref{fig:nArcs}. Again, the green shaded area shows the observational result.}
\label{fig:nArcs2}
\end{figure}

Figure~\ref{fig:nArcs2} shows the number of arcs with length-to-width ratio $d \ge 10$ predicted to be observed in a $\Lambda$CDM cosmological model as a function of $\sigma_8$, accounting for cluster mergers, with and without the inclusion of the scatter in the halo concentration. To keep the plot readable, we only show results for a limiting magnitude of $21.5$ in both the I and R bands. The scatter in the mass-concentration relation increases the number of arcs by $\sim 70\%$ for $\sigma_8 = 0.7$, and only of $\sim 20\%$ for $\sigma_8 = 0.9$. This decrease is due to the fact that the concentration increases with $\sigma_8$, reducing the relative effect of fluctuations around the nominal value. On the whole, the scatter in the halo concentrations does not help much in improving the agreement with observations: models with high $\sigma_8$ still fall short by a factor $\sim 2$, and models with low $\sigma_8$ are still an order of magnitude off.

\section{Summary and Discussion}\label{sct:con}

We have computed the optical depth, differential optical depth
and total number of gravitational arcs with length-to-width ratios $d\ge7.5$ and $d\ge10$ expected in model universes with five different normalisations $\sigma_8$. 
The values for the matter and dark-energy density parameters as well as for the Hubble constant were taken from the 3-year data release of the WMAP satellite, while $\sigma_8$ is taken from the set $\left\lbrace 0.7,0.75,0.8,0.85,0.9 \right\rbrace$.

The cluster population is modelled planting Monte-Carlo merger trees for a set of $\mathcal{N}=1,000$ dark-matter halos, uniformly drawn at $z=0$ from the mass range $\left[10^{14},2.5 \times 10^{15}\right] M_\odot h^{-1}$. Each halo is assumed to have an NFW density profile and elliptically distorted lensing potentials, with ellipticity $\epsilon = 0.3$. The effect on the strong lensing efficiency of the interaction with substructures is also taken into account as described in Sect.~\ref{sct:clu}. Background sources are properly distributed in redshift following to the observed distribution of equation~(\ref{eqn:zs}).

We converted the average optical depth into an observed number of gravitational arcs using an appropriate flux distribution function for background sources, taking the magnification bias into account. We considered three different limiting magnitudes in both the I and R bands, including the R-band limit set by observational studies \citep{GI94.1,LU99.1}. 

Confirming straightforward expectations, we find that the total strong-lensing efficiency grows steeply with $\sigma_8$. We find that the number of arcs observable in a cosmological model with $\sigma_8 = 0.7$ is up to one order of magnitude below the number of arcs when $\sigma_8 = 0.9$.

The effect of cluster mergers also depends on cosmology. Mergers with relatively small substructures are more likely in a low-$\sigma_8$ universe, in which individual galaxy clusters are also less efficient lenses because of their lower concentration, thus causing the effect of mergers to become more pronounced. Cluster mergers increase the total optical depth by up to a factor of $\sim 5$ in a model with $\sigma_8 = 0.7$ and by a factor of $\sim 2$ to $3$ in a model with $\sigma_8 = 0.9$.

A particularly strong effect is that the differential optical depth at $z\gtrsim1$ for low $\sigma_8$ can be up to several orders of magnitude smaller than for high $\sigma_8$. This causes a severe problem for explaining the high observed incidence of large arcs in high redshift clusters if $\sigma_8$ is as low as inferred from the WMAP-3 data. Models with low $\sigma_8$ also significantly fail to reproduce the number of arcs observed in complete, X-ray selected galaxy clusters samples. For high $\sigma_8\gtrsim0.9$, agreement with the observations can be achieved. Thus, we conclude that the arc statistics problem is unsolved based on the WMAP-3 parameters even if a suitable source redshift distribution is included and cluster mergers are taken into account.

We also included the scatter in the relation between halo mass and concentration in our calculation, similarly to \cite{FE07.3}. This increases the total lensing efficiency only slightly, by about $20\%$ for high $\sigma_8$ and by $\sim 70\%$ for low $\sigma_8$. An increment by a factor of $\sim 2$ may be contributed by the inclusion of finer details of the cluster structure, like the dark matter halos of single member galaxies \citep{ME00.1,FL00.1}. The order of magnitude discrepancy revealed by models with low $\sigma_8$ would however persist.

We mention that, as we verified, increasing the matter density parameter above the fiducial value $\Omega_{\mathrm{m},0} = 0.265$ used here, and keeping the flat universe assumption, increases the abundance of observed giant arcs. This is due to the fact that a higher $\Omega_{\mathrm{m},0}$ implies a larger cluster abundance today, and even though the evolution of the cluster population is faster due to the minor contribution of the cosmological constant, this is not enough to counteract the effect in the redshift range important for strong lensing. For instance, a WMAP-1 cosmology with $\sigma_8 = 0.9$ and $\Omega_{\mathrm{m},0} = 0.3$ increases the number of arcs with respect to the values reported in Figure \ref{fig:nArcs} for $\sigma_8 = 0.9$, but only of $\sim 60\%$. On the other hand, if $\sigma_8 = 0.8$ is fixed, the model results can be brought in agreement with the observations only with unrealistically high values of $\Omega_{\mathrm{m},0} \gtrsim 0.4$.
A study of the effect on arcs statistics of the combined variation of $\sigma_8$ and $\Omega_{\mathrm{m},0}$, for instance following the degeneracy direction given by some cosmological test, is beyond the scope of this work, but is certainly interesting for future analysis.

\begin{figure}[t]
  \includegraphics[width=\hsize]{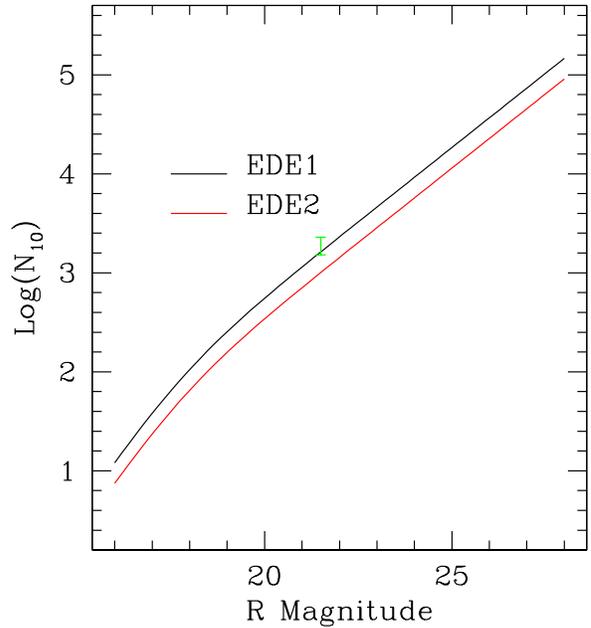}
\caption{The number of arcs predicted in two examples of cosmological models with early-dark energy, chosen such as to agree with CMB, supernovae and large-scale structure data. In such models, observational constraints (shown here as the green vertical bar) are easily met even with low $\sigma_8$. Additional details on the EDE models presented here can be found in \cite{FE07.1}}
\label{fig:nArcsEDE}
\end{figure}

Some comments regarding the choice of our general assumption that dark-matter halos have NFW density profiles may be in order. Recent publications \citep{NA04.1,PR06.1,ME06.1,GA07.1} have pointed out that the NFW profile is indeed not the best possible representation of dark-matter halo profiles found in high-resolution $N$-body simulations. This is mainly because the concentrations found depend on the actual radial range used for profile fitting. According to \cite{ME06.1}, the \cite{EI65.1} profile yields better fits, originally proposed to model the distribution of stars in the Milky Way. The same authors show that the original NFW prescription for relating the concentration and the virial mass of a dark matter halo works better than the subsequent extensions by \cite{BU01.1} and \cite{EK01.1}, even though it must be recalled that they refer to the concentrations relative to the scale radius of the Einasto profile. However, the \cite{EK01.1} prescription used here is still a reasonable fit to the mass-concentration relation shown by \cite{GA07.1} at $z \sim 1$, and since there are only a handful of efficient clusters at redshifts significantly larger than unity even in models with high $\sigma_8$, we believe this to be a sound approximation.

There are two possible ways of looking at the consequence of the results obtained here. Either, the normalisation is high, $\sigma_8 \gtrsim 0.9$, and the low value derived from the WMAP data is spurious because of some problem hidden in the data-reduction process or in the Galactic foregrounds \citep{DE04.1,CR06.1}. Or, and most likely, the actual $\sigma_8$ is in fact low, as recent weak-lensing measurements seem to confirm \citep{FU08.1}, but we fail to properly understand some part of galaxy-cluster physics affecting both the number counts and the relative strong-lensing properties of galaxy clusters.

After the submission of this paper, the five-year data release of the WMAP satellite was made available. There, the value of the matter density parameter slightly increased with respect to the WMAP-3 release, to the best fit $\Omega_{\mathrm{m},0} = 0.279$ (when CMB data are combined with distance measures from Ia-type supernovae and baryon acoustic oscillations). The power-spectrum normalization also increased to $\sigma_8 = 0.817$, while the error bars decreased (0.026 at $68\%$ confidence level for WMAP-5). As mentioned above, we expect this slight increase in the matter density parameter to be insignificant for our results. On the other hand, with $\sigma_8 \sim 0.8$, the discrepancy between the model predictions and the observations is reduced to a factor of $>6$. This is less than the discrepancy with WMAP-3 data, but still cannot be accommodated by known contributions. For instance, different kinds of baryonic physics can increase the cluster cross sections  of $\lesssim 100\%$. In addition, the arc abundance observed in an X-ray selected cluster sample is only a fraction of the total one \citep{FE07.2}. This effect is not included in the present work, and is likely to worsen the agreement with theoretical studies.

Regarding the abundance of arcs in high-redshift clusters (see Figure \ref{fig:cumulative}), a value of $\sigma_8 \sim 0.8$ gives many more distant arcs compared to a model with $\sigma_8 = 0.7 - 0.75$, even though a factor of $\sim 3$ discrepancy compared to high-normalisation models is still present. Hence, the high incidence of distant arcs remains a problem for a WMAP-5 cosmology, even more than it was for a high-normalisation WMAP-1 model. 

In conclusion, the newest WMAP data do not remove the discrepancy between theory and observations for the statistics of giant arcs, even though it is somewhat reduced with the new parameter values.

A way out may be possible in presence of a dynamical dark energy component, cf.~Figure~\ref{fig:nArcsEDE}. As shown in \cite{FE07.1}, the presence of early-dark energy (EDE briefly) can play the role of an increased $\sigma_8$ on non-linear scales because it tends to shift the entire structure-formation process to earlier times \citep{BA06.1}. Consequently, the production of gravitational arcs, and in particular the lensing efficiency for high-redshift clusters, are significantly increased. Alternatively, non-Gaussian density fluctuations may have similarly strong and positive effects (\citealt{GR07.1}; See also \citealt{MA04.2,SA07.1}). Future studies directed at the recovery of a possible redshift evolution of the dark-energy density will be fundamentally important also in this context \citep{BA03.1,ME05.1,ME05.2}. Forthcoming analyses of the arc statistics problem also require a substantial increase of the observational data basis, which will be enabled by large-scale optical or infrared surveys in conjunction with fast algorithms for automatic arc detection \citep{LE04.1,HO05.1,CA07.1,SE07.1}.

\section*{Acknowledgements}

We acknowledge financial contributions from contracts ASI-INAF I/023/05/0, ASI-INAF I/088/06/0 and INFN PD51. This work was supported in part by the Sonderforschungsbereich SFB 439 of the Deutsche Forschungsgemeinschaft. We wish to thank the anonymous referee for useful remarks that allowed us to improve the presentation of our work.

\bibliographystyle{aa}
\bibliography{./master}

\end{document}